\documentclass[twocolumn,showpacs,preprintnumbers,superscriptaddress,amsmath,floatfix,amssymb,prd]{revtex4}
\usepackage[colorlinks=true]{hyperref}
\usepackage{graphicx}

\newcommand{\comment}[1]{}

\newcommand{\lr}[1]{ \left( #1 \right) }
\newcommand{\lrs}[1]{ \left[ #1 \right] }
\newcommand{\lrc}[1]{ \left\{ #1 \right\} }
\newcommand{\vev}[1]{ \langle \, #1 \, \rangle }

\newcommand{\tr}{ {\rm Tr} \, }

\newcommand{\ket}[1]{ \, | #1 \rangle }
\newcommand{\bra}[1]{ \langle #1 | \, }
\newcommand{\braket}[2]{\langle #1 | #2 \rangle}

\newcommand{\const}{ {\rm const}}

\renewcommand{\th}{ {\rm th} \,  }

\newcommand{\expa}[1]{ \exp{\left( #1 \right)} }

\newcommand{\logo}{\\ \vskip -18mm
\leftline{\includegraphics[scale=0.3,clip=false]{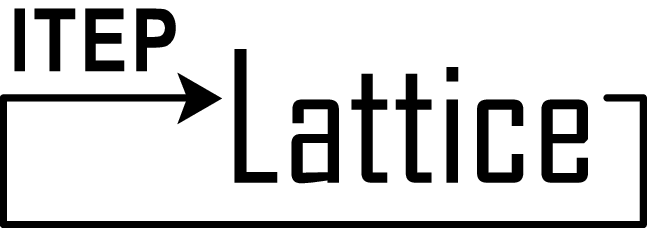}} \vskip 10mm}

\begin{document}
\sloppy
\preprint{ITEP-LAT/2008-14}

\title{Entanglement entropy in gauge theories and the holographic principle for electric strings\logo}
\author{P. V. Buividovich}
\email{buividovich@tut.by}
\affiliation{JIPNR, National Academy of Science, 220109 Belarus, Minsk, Acad. Krasin str. 99}
\affiliation{ITEP, 117218 Russia, Moscow, B. Cheremushkinskaya str. 25}
\author{M. I. Polikarpov}
\email{polykarp@itep.ru}
\affiliation{ITEP, 117218 Russia, Moscow, B. Cheremushkinskaya str. 25}
\date{June 20, 2008}
\begin{abstract}
 We consider quantum entanglement between gauge fields in some region of space $A$ and its complement $B$. It is argued that the Hilbert space of physical states of gauge theories cannot be decomposed into a direct product $\mathcal{H}_{A} \otimes \mathcal{H}_{B}$ of Hilbert spaces of states localized in $A$ and $B$. The reason is that elementary excitations in gauge theories - electric strings - are associated with closed loops rather than points in space, and there are closed loops which belong both to $A$ and $B$. Direct product structure and hence the reduction procedure with respect to the fields in $B$ can only be defined if the Hilbert space of physical states is extended by including the states of  electric strings which can open on the boundary of $A$. The positions of string endpoints on this boundary are the additional degrees of freedom which also contribute to the entanglement entropy. We explicitly demonstrate this for the three-dimensional $Z_{2}$ lattice gauge theory both numerically and using a simple trial ground state wave function. The entanglement entropy appears to be saturated almost completely by the entropy of string endpoints, thus reminding of a ``holographic principle'' in quantum gravity and AdS/CFT correspondence.
\end{abstract}
\maketitle

 Geometric entanglement entropy \cite{tHooft:85:1, Calabrese:04:1, Calabrese:06:1} of confining gauge theories has recently become a subject of extensive studies, mostly due to the discovery of its non-analytical behavior with respect to the size of the region. This non-analyticity was first predicted in the framework of AdS/CFT correspondence in \cite{Takayanagi:06:1, Klebanov:07:1}, and then within the Migdal-Kadanoff approximation in lattice gauge theories \cite{Velytsky:08:1}. Finally, a signature of nonanalytic behavior of entanglement entropy was found in numerical simulations of $SU\lr{2}$ lattice gauge theory \cite{Buividovich:08:2}.

 In order to define the geometric entanglement entropy of some region $A$, one should represent the Hilbert space $\mathcal{H}$ of the field theory under consideration as a direct product of Hilbert spaces $\mathcal{H}_{A}$ and $\mathcal{H}_{B}$ of states localized in $A$ and its complement $B$ \cite{Bernstein:96:1}: $\mathcal{H} = \mathcal{H}_{A} \otimes \mathcal{H}_{B}$. For instance, for scalar field theory the spaces $\mathcal{H}_{A}$ and $\mathcal{H}_{B}$ can be constructed by acting on some proper initial state with field operators localized within $A$ and $B$. In other words, elementary excitations of scalar fields are associated with points in space, and any point can be classified as belonging either to $A$ or $B$ (except for the set of points on the boundary, which has zero measure). One then defines the reduced density matrix for the fields in $A$ as a partial trace of the density matrix of the ground state $\ket{0}$ of the theory over $\mathcal{H}_{B}$: $\hat{\rho}_{A} = \tr_{B} \ket{0} \bra{0}$. Geometric entanglement entropy $S\lrs{A}$ is the von Neumann entropy of the reduced density matrix $\hat{\rho}_{A}$ \cite{Bernstein:96:1}:
\begin{eqnarray}
\label{EE_def}
S\lrs{A} = - \tr_{A} \hat{\rho}_{A} \ln \hat{\rho}_{A}
\end{eqnarray}

 However, in none of the works \cite{Takayanagi:06:1, Klebanov:07:1, Velytsky:08:1, Buividovich:08:2} such a direct product structure was explicitly constructed for gauge theories or even used. The papers \cite{Takayanagi:06:1, Klebanov:07:1} mainly rely on the conjecture of \cite{Ryu:06:1} which relates the entanglement entropy of conformal field theories living on the boundary of AdS space with certain minimal hypersurfaces in the bulk of this space. The results of \cite{Velytsky:08:1, Buividovich:08:2} are based on the relation between entanglement entropy and the free energies of the theory on the spaces with multiple cuts, which was originally derived in \cite{Calabrese:04:1, Calabrese:06:1} for scalar field theories.

 The aim of the present work is to explicitly construct a decomposition of the Hilbert space of states of gauge theories into a direct product $\mathcal{H}_{A} \otimes \mathcal{H}_{B}$. One of our main assertions is that such a decomposition is impossible for a Hilbert space of physical states, i.e. a Hilbert space of states which satisfy the Gauss law. The reason is that in contrast to scalar field theories, in pure gauge theories elementary excitations are not associated with points in space, but rather with closed loops, which are the lines of electric flux or electric strings \cite{PolyakovGaugeStrings, Kogut:75:1, Kogut:79:1}. Correspondingly, closed loops cannot be classified as belonging either to $A$ or $B$, since the set of loops which belong to both $A$ and $B$ has nonzero measure in the space of all loops.

\begin{figure}
  \includegraphics[width=6cm]{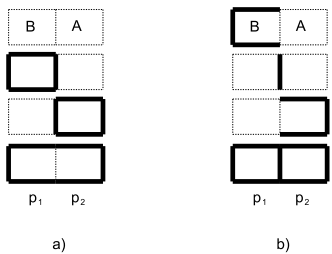}\\
  \caption{Basis quantum states for the simplest two-dimensional lattice. Thick lines are the links with $m_{l} = 1$.}
  \label{fig:simplest_lattices}
\end{figure}

 For the sake of brevity let us consider the simplest case of $Z_{2}$ lattice gauge theory in $\lr{2 + 1}$ space-time dimensions. The arguments presented below can be generalized to other gauge theories in a straightforward way. The Hamiltonian of this theory is:
\begin{eqnarray}
\label{z2hamiltonian}
\hat{H} = \frac{1}{2 \beta_{g}}\, \sum \limits_{l} \hat{\sigma}_{l} - \beta_{g} \sum \limits_{p} \prod \limits_{l \in p} \hat{z}_{l}
\end{eqnarray}
where $\beta_{g}$ is the coupling constant, $\sum \limits_{l}$ and $\sum \limits_{p}$ denote summation over all lattice links and plaquettes and $\hat{z}_{l}$ and $\hat{\sigma}_{l}$ are the operators with eigenvalues equal to $\pm 1$ which satisfy $\lrc{\hat{\sigma}_{l},\hat{z}_{l}} = 0$. The Hilbert space is a space of all functions $\psi\lrs{ z_{l} }$ of $Z_{2}$-valued link variables $z_{l}$ on a two-dimensional lattice. A convenient basis in this space can be parameterized by a set of labels $m_{l} = 0, 1$ defined on each link of the lattice: $\psi\lrs{z_{l}; m_{l}} \sim \prod \limits_{l} z_{l}^{m_{l}}$. This basis is orthonormal with respect to the scalar product $\braket{\psi_{1}}{\psi_{2}} = \sum \limits_{\lrc{z_{l}}} \bar{\psi}_{1}\lrs{z_{l}} \psi_{2}\lrs{z_{l}}$. The Hilbert space of physical states $\mathcal{H}_{0}$ is obtained by imposing the Gauss law as a first-class constraint, which leaves only the wave functions which are invariant under gauge transformations: $\psi\lrs{ z_{l}^{\Omega} } = \psi\lrs{z_{l}}$. The basis states in $\mathcal{H}_{0}$ can be labelled by such sets $\lrc{m_{l}}$, for which the links with $m_{l} = 1$ form closed non-self-intersecting loops, and can be obtained by acting with all possible Wilson loop operators $W\lrs{C} = \prod \limits_{l \in C} z_{l}$ on the trivial strong-coupling ground state with $\psi\lrs{z_{l}} = \const$. The operator $W\lrs{C}$ thus creates an electric string on the loop $C$.

 As the simplest example consider the lattice which consists of just two plaquettes, $p_{1}$ and $p_{2}$ (see Fig. \ref{fig:simplest_lattices}). The Hilbert space of physical states on such lattice is four-dimensional and contains the states with electric flux flowing either around plaquettes $p_{1}$ or $p_{2}$, or around both: $\psi_{1}\lrs{z_{l}} = 1$, $\psi_{2}\lrs{z_{l}} = z_{p_{1}}$, $\psi_{3}\lrs{z_{l}} = z_{p_{2}}$, $\psi_{4}\lrs{z_{l}} = z_{p_{1}} \, z_{p_{2}}$, where $z_{p} = \prod \limits_{l \in p} z_{l}$. By direct counting of states one can make sure that the Hilbert space spanned on this four states can not be represented as a direct product of Hilbert spaces of functions on any complementary sets of links.

  A proof of this assertion in the general case is also rather obvious. Let us denote the Hilbert spaces of all functions on links within $A$ or $B$ as $\tilde{\mathcal{H}}_{A}$ and $\tilde{\mathcal{H}}_{B}$. We would like to decompose the Hilbert space of physical states $\mathcal{H}_{0}$ as $\mathcal{H}_{0} = \mathcal{H}_{A} \otimes \mathcal{H}_{B}$, so that $\mathcal{H}_{A}$ and $\mathcal{H}_{B}$ are the subspaces of $\tilde{\mathcal{H}}_{A}$ and $\tilde{\mathcal{H}}_{B}$.  Consider some state with the wave function of the form $\psi\lrs{z_{l}} \sim \prod \limits_{C} z_{l}$, where the $C$ is some closed loop which belongs both to $A$ and $B$. This state is a direct product of two states in $\tilde{\mathcal{H}}_{A}$ and $\tilde{\mathcal{H}}_{B}$ with wave functions $\psi_{A}\lrs{z_{l}} \sim \prod \limits_{C_{A}} z_{l}$ and $\psi_{B}\lrs{z_{l}} \sim \prod \limits_{C_{B}} z_{l}$, where $C_{A}$ and $C_{B}$ are the parts of the loop $C$ which belong to $A$ and $B$. Since the basis states corresponding to different loops are orthogonal, this decomposition is unique, and the states $\psi_{A}$ and $\psi_{B}$ should necessarily belong to $\mathcal{H}_{A}$ and $\mathcal{H}_{B}$. The spaces $\mathcal{H}_{0}$ and hence $\mathcal{H}_{A}$ and $\mathcal{H}_{B}$ should also contain the trivial strong-coupling ground state with $\psi_{0}\lrs{z_{l}} = \const = \psi_{A \, 0} \psi_{B \, 0}$, $\psi_{A, B \, 0}\lrs{z_{l}} = \const$. Therefore if $\mathcal{H}_{0}$ is indeed a direct product of $\mathcal{H}_{A}$ and $\mathcal{H}_{B}$, it should also contain a direct product of, say, $\psi_{A \, 0}$ with $\psi_{B}$. The wave function for such a state is proportional to $\prod \limits_{C_{B}} z_{l}$, but $C_{B}$ is in general not closed, therefore the Gauss law is violated at the boundary between $A$ and $B$. We thus arrive at the contradiction, which completes the proof.

 Let us now try to extend the Hilbert space of physical states in some minimal way, so that a direct product structure can be introduced. Consider again the lattice on Fig. \ref{fig:simplest_lattices}. Since the cause of the trouble are the strings which cross the boundary between $A$ and $B$ (this is the link in the middle on Fig. \ref{fig:simplest_lattices}), let us try to cut such strings along this boundary. This yields four more states, which are schematically shown on Fig. \ref{fig:simplest_lattices}, b. The resulting extended Hilbert space is now eight-dimensional. Assume that the link in the middle and all the links to the right of it belong to the region $A$, and that all other links belong to the region $B$. It is easy to see that the extended Hilbert space can be represented now as a direct product of a two-dimensional Hilbert space of functions on the links in $B$ times a four-dimensional Hilbert space of functions on the links in $A$, which are not invariant under the gauge transformations only at the endpoints of the link in the middle. In the general case, the minimal extension $\tilde{\mathcal{H}}_{0}$ of $\mathcal{H}_{0}$ is also the Hilbert space of states which violate the Gauss law only at the boundary $\partial A$ between $A$ and $B$. Minimal extension here means that the Gauss law does not hold in a minimal number of points, and $\tilde{\mathcal{H}}_{0}$ is a minimal subspace of $\mathcal{H}$ which contains $\mathcal{H}_{0}$ and which can be represented as a direct product. To see that this is indeed so, suppose that the Gauss law still holds in some points on $\partial A$. In this case one can repeat all the arguments above for the loop $C$ which goes through these points and arrive at the same contradiction. It should be mentioned that such an extension of Hilbert space of physical states was already implicitly used in the study of topological entanglement entropy \cite{Hamma:05:1, Levin:06:1} and entanglement entropy of spin networks \cite{Donnelly:08:1}, although only for a particular class of ground states which can be thought of as the ground states of lattice gauge theories in the limit of infinitely strong coupling. In the extended Hilbert space one can naturally define the partial trace over the fields in $B$ as a sum over all links which belong to $B$:
\begin{eqnarray}
\label{reduction}
\bra{z_{l}'} \, \tr_{B} \hat{\rho} \, \ket{z_{l}''} = \sum \limits_{\lrc{z_{l}}, l \in B} \rho\lrs{ \lrc{z_{l}, z_{l}'}, \lrc{z_{l}, z_{l}''}}
\end{eqnarray}
If the expression (\ref{reduction}) was simply postulated without introducing the extended Hilbert space, it would lead to peculiar density matrices which satisfy the Gauss law, i.e., which commute with the generators of gauge transformations, but which nevertheless can not be represented in the form $\hat{\rho}_{A} = \sum \limits_{i} w_{i} \ket{\psi_{i}} \bra{\psi_{i}}$, where $\ket{\psi_{i}}$ are physical states. As an example, consider again the state with the wave function $\psi\lrs{z_{l}} \sim \prod \limits_{l \in C} z_{l}$, where $C$ is some loop which belongs both to $A$ and $B$. The reduced density matrix $\rho_{A}\lrs{z_{l}', z_{l}''} \sim \prod \limits_{l \in C_{A}} z_{l}' z_{l}''$ is invariant under simultaneous gauge transformations of $z_{l}'$ and $z_{l}''$, but obviously can not be decomposed over the basis of physical states. On the other hand, in the extended Hilbert space such decomposition is trivial. Thus one could in principle arrive at the concept of the extended Hilbert space by trying to interpret the outcome of the reduction procedure (\ref{reduction}).

\begin{figure}
  \includegraphics[width=6cm]{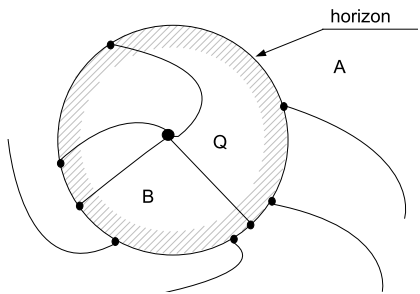}\\
  \caption{Thin tubes of electric flux in the vicinity of a charged black hole with charge $Q = 4$.}
  \label{fig:black_hole_hair}
\end{figure}

 We thus conclude that in order to define the spaces of states of gauge fields which are localized either in $A$ or $B$ one should extend the Hilbert space of physical states and include also the states of electric strings which can open on the boundary $\partial A$ between $A$ and $B$. This boundary can be therefore considered as a sort of $D$-brane for electric strings, with the positions of their endpoints being the additional degrees of freedom which emerge as a result of such extension. It is reasonable to conjecture that these extra degrees of freedom should contribute somehow to the entropy of entanglement between $A$ and $B$. In order to understand how this can happen, it is instructive to consider a confining Abelian gauge theory, such as compact QED in the strong coupling limit \cite{PolyakovGaugeStrings}, which interacts with a charged black hole with charge $Q$ (see Fig. \ref{fig:black_hole_hair}). Since the region inside the horizon is not accessible to an external observer, in order to find the density matrix of the fields outside of the black hole one should trace over all states of fields inside the horizon, which is one possible way to relate the Bekenstein-Hawking entropy with  the entanglement entropy \cite{Bombelli:86:1, Fursaev:06:1}. For confining theory the charge should be connected with the horizon with $|Q|$ thin tubes of electric field, and $|Q|$ such tubes should also go out of a black hole. The positions of the endpoints of these flux tubes outside and inside of the horizon should be completely uncorrelated, otherwise such a system would violate the ``no - hair'' theorem. On the other hand, before the formation of a black hole all flux tubes were continuous, and ended only on the charge $Q$. The entropy associated with the uncertainty in the positions of endpoints on the horizon should therefore also contribute to the entropy of a black hole and hence to the entanglement entropy of the fields behind the horizon.

 In order to estimate the contribution of these additional degrees of freedom on the boundary $\partial A$, let us first calculate the entanglement entropy for a simple trial ground state wave function of $Z_{2}$ lattice gauge theory in $\lr{2 + 1}$ space-time dimensions. This trial ground state is a superposition of all possible configurations of closed electric strings with the weight $\expa{ - \frac{\alpha}{2} \, \sum \limits_{l} m_{l}}$, where $\sum \limits_{l} m_{l}$ is the total length of strings:
\begin{eqnarray}
\label{trial_wf}
\Psi_{0}\lrs{z_{l}} = C \sum \limits_{\lrc{ \delta m_{l} = 0}}
\expa{ - \frac{\alpha}{2} \, \sum \limits_{l} m_{l}} \psi\lrs{z_{l}; m_{l}}
\end{eqnarray}
where $\delta m_{l} = 0$ means that we sum only over closed electric strings and $C$ is the normalization constant. Fortunately, the trial ground state (\ref{trial_wf}) yields an expression for the entanglement entropy which has a universal form, and which, as we will see from the results of numerical simulations, seems to be a good approximation for the entanglement entropy of a true ground state of $Z_{2}$ lattice gauge theory for all values of the coupling constant. The wave function (\ref{trial_wf}) would be a product of functions of individual link variables if the constraint $\delta m_{l} = 0$ was omitted. Thus in some sense the trial ground state (\ref{trial_wf}) is only ``minimally'' entangled because the electric strings are closed, and there is no entanglement between closed string states in the bulk of the regions $A$ and $B$. Despite its simplicity, the wave function (\ref{trial_wf}) can qualitatively describe the most essential features of the dynamics of electric strings in $Z_{2}$ lattice gauge theory, such as the emergence of percolating strings in the vicinity of a quantum phase transition, due to the dimensional reduction property \cite{Greensite:79:1}.

 The entanglement entropy for the state (\ref{trial_wf}) can be calculated using the ``replica trick'' $S\lrs{A} = - \lim \limits_{s \rightarrow 1} \, \frac{\partial}{\partial s} \, \tr \hat{\rho}_{A}^{s}$ \cite{Calabrese:04:1, Calabrese:06:1}. Applying the reduction procedure (\ref{reduction}) and using the orthogonality of the basis states $\psi\lrs{z_{l}; m_{l}}$, one can rewrite the trace $\tr \hat{\rho}_{A}^{s}$ as:
\begin{eqnarray}
\label{trial_wf_reduced_dm_trace1}
\tr \hat{\rho}_{A}^{s} = C^{2 s}
\sum \limits_{\lrc{m_{l\,1}}} \ldots \sum \limits_{\lrc{m_{l\,s}}}
\expa{ -\alpha \sum \limits_{k = 1}^{s} \sum \limits_{l} m_{l \, k} }
\end{eqnarray}
Here one should sum only over such $\lrc{m_{l\,k}}$ for which the electric strings in $A$ described by $\lrc{m_{l\,k}}$, $l \in A$ and $\lrc{m_{l\,k+1}}$, $l \in A$ end in the same points on $\partial A$ as the string in $B$ described by $\lrc{m_{l\,k}}$, $l \in B$. Iterating this constraint for all $k = 1, \ldots, s$, we see that we actually sum over $s$ independent string configurations in $A$ and $B$, which all intersect $\partial A$ in the same points $x_{1}, \ldots, x_{m}$. It is convenient to sum separately over the coordinates of these endpoints and over all configurations of electric strings in $A$ and $B$ which end in $x_{1}, \ldots, x_{m}$. After such rearrangement the sums over $\lrc{m_{l\,k}}$, $l \in A$ and $\lrc{m_{l\,k}}$, $l \in B$ in (\ref{trial_wf_reduced_dm_trace1}) factorize:
\begin{eqnarray}
\label{trial_wf_reduced_dm_trace2}
\tr \hat{\rho}_{A}^{s} = \sum \limits_{\lrc{x_{1}, \ldots, x_{m}}}
\lr{
\sum' \limits_{\lrc{m_{l}}} C^{2}\, \expa{ - \alpha \sum \limits_{l} m_{l} }
}^{s}
\end{eqnarray}
where the prime over the sum in the brackets means that that the electric strings described by $\lrc{m_{l}}$ should intersect $\partial A$ only in the points $\lrc{x_{1}, \ldots, x_{m}}$. Remember now that $\expa{ - \alpha \sum \limits_{l} m_{l} }$ is the square of the trial wave function (\ref{trial_wf}) in the basis of states $\psi\lrs{z_{l}; m_{l}}$, or, in other words, simply the probability distribution of the configurations of electric strings on the lattice. The constrained sum $\sum' \limits_{\lrc{m_{l}}} C^{2}\, \expa{ - \alpha \sum \limits_{l} m_{l} }$ is then nothing but the probability distribution $p\lrs{\lrc{x_{1}, \ldots, x_{m}}}$ of the intersection points between electric strings and the boundary $\partial A$ in the ground state of the theory. Taking now the derivative of (\ref{trial_wf_reduced_dm_trace2}) at $s = 1$ and using the replica trick, we find that for the trial ground state (\ref{trial_wf}) the entanglement entropy is simply the classical Shannon entropy of the probability distribution $p\lrs{\lrc{x_{1}, \ldots, x_{m}}}$:
\begin{eqnarray}
\label{trial_wf_ent_ent}
S\lrs{A} = - \sum \limits_{\lrc{x_{1}, \ldots, x_{m}}} p\lrs{\lrc{x_{1}, \ldots, x_{m}}} \ln p\lrs{\lrc{x_{1}, \ldots, x_{m}}}
\end{eqnarray}
Although this result was derived only for a particular wave function (\ref{trial_wf}), its simple and universal form suggests that it can be applied to any quantum state, for which the entanglement between the states of the closed strings in the bulk of the regions $A$ and $B$ can be neglected. An expression similar to (\ref{trial_wf_ent_ent}) has been also obtained recently for the entanglement entropy in loop quantum gravity \cite{Donnelly:08:1}.

 For the purposes of numerical simulations, it is convenient to represent the expression (\ref{trial_wf_ent_ent}) in a somewhat different form, assuming for simplicity that for a given $m$ all the intersection points $x_{1}, \ldots, x_{m}$ are not correlated. Taking into account that only one electric string can pass through one lattice site, after some simple transformations one can rewrite the expression (\ref{trial_wf_ent_ent}) as:
\begin{eqnarray}
\label{entropy_simplification}
S\lrs{A} = S_{0}\lrs{ P\lr{m} } +  \vev{\ln\lr{\frac{|\partial A|!}{m! \, \lr{|\partial A| - m}!}}}
\end{eqnarray}
where $S_{0}\lrs{ P\lr{m} }$ is the entropy of the probability distribution $P\lr{m}$ of the number of intersection points and $|\partial A|$ is the total length of the boundary $\partial A$ in lattice units.

\begin{figure}
  \includegraphics[width=8cm]{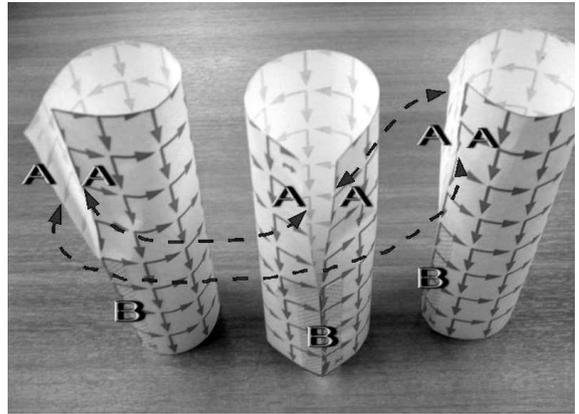}\\
  \caption{Gluing of $s = 3$ $\lr{1 + 1}$-dimensional square lattices into a three-sheeted Riemmann surface.}
  \label{fig:gluing}
\end{figure}

 In order to check the validity of the expression (\ref{trial_wf_ent_ent}) for the true ground state of $Z_{2}$ lattice gauge theory, we have measured the entanglement entropy and the entropy of the probability distribution of intersection points $p\lrs{\lrc{x_{1}, \ldots, x_{m}}}$ in lattice Monte-Carlo simulations. The region $A$ was a square of size $l\times l$. To calculate the entanglement entropy, we have used the same method and the same approximations as in the work \cite{Buividovich:08:2}, measuring only the differences of entropies $S\lrs{A'} - S\lrs{A} = S\lr{l+1} - S\lr{l}$. Let us note here that the arguments presented in this Letter actually justify the method used in \cite{Buividovich:08:2, Velytsky:08:1} to calculate the entanglement entropy of lattice gauge theories. The lattices with cuts used in \cite{Buividovich:08:2, Velytsky:08:1} have the topology $\mathbb{C}_{\lr{s}} \otimes \mathbb{T}^{D-2}$ \cite{Calabrese:04:1, Calabrese:06:1}, where $\mathbb{C}_{\lr{s}}$ is the $s$-sheeted Riemann surface, $\mathbb{T}^{D-2}$ is the $\lr{D-2}$-dimensional torus and $D$ is the dimensionality of space-time, and were constructed by gluing together $s$ square lattices along the links which belong to the region $A$ (see Fig. \ref{fig:gluing}). Such construction essentially relies on the notion of the partial trace (\ref{reduction}) over the links in $B$ , which only makes sense for the extended Hilbert space constructed above (see \cite{Calabrese:04:1, Calabrese:06:1} and Appendix A in \cite{Buividovich:08:2} for a detailed derivation).

\begin{figure}
  \includegraphics[width=6cm, angle=-90]{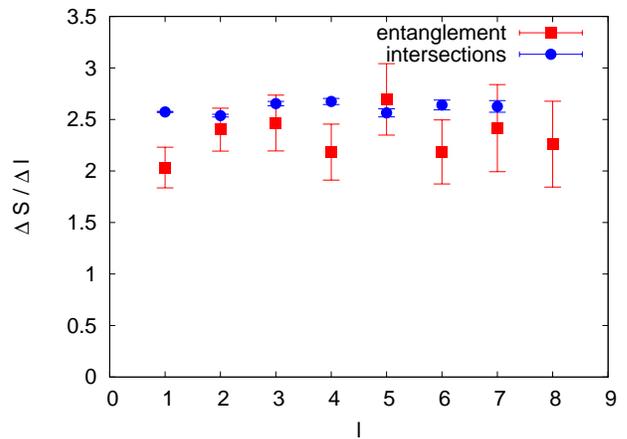}\\
  \caption{Lattice derivatives of the entanglement entropy and the entropy of intersection points over the size of the region $A$ at $\beta_{g} = 0.788$, $\beta_{s} = 0.21$, $16^{3}$ lattice.}
  \label{fig:ment_vs_l}
\end{figure}

\begin{figure}
  \includegraphics[width=6cm, angle=-90]{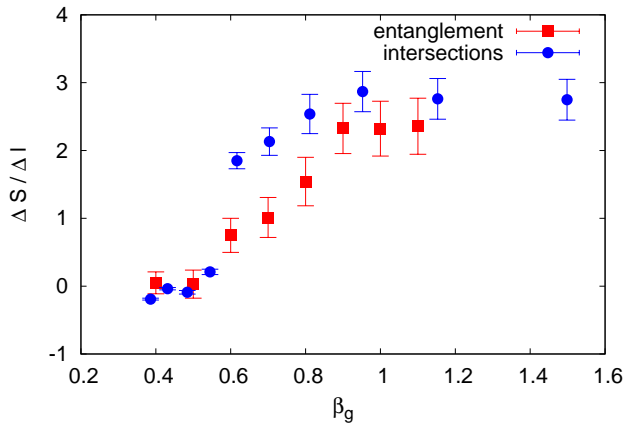}\\
  \caption{Lattice derivatives of the entanglement entropy and the entropy of intersection points over the size of the region at $l = 8$ at different values of the coupling constant $\beta_{g}$ for $16^{3}$ lattice.}
  \label{fig:ment_vs_beta}
\end{figure}

 Similarly to the entanglement entropy, the positions of intersection points between electric strings and the boundary $\partial A$ are not directly measurable in Monte-Carlo lattice simulations. The reason is that the corresponding operator is not diagonal in the eigenbasis of the link variable operator $\hat{z}_{l}$. Fortunately, for $Z_{2}$ lattice gauge theory in $\lr{2 + 1}$ dimensions one can use the Kramers-Wannier duality transformation and express the expectation values of field operators which are diagonal in the basis $\psi\lrs{z_{l}; m_{l}}$ as expectation values in the dual theory \cite{Kogut:79:1, PolyakovGaugeStrings}, which is the Ising model in $\lr{2 + 1}$ dimensions with the Hamiltonian $\hat{H} = \frac{1}{2 \beta_{s}}\, \sum \limits_{s} \hat{\sigma}_{s} - \beta_{s} \sum \limits_{l} \hat{z}_{s} \hat{z}_{s+l} $, where $\sum \limits_{s}$ denotes the sum over all lattice sites. The configurations of electric strings in the gauge theory can be identified with the closed loops formed by the links which are dual to the links with $z_{s} z_{s+l} = - 1$ in the Ising model, if the coupling constant $\beta_{s}$ is related to the coupling constant $\beta_{g}$ of the gauge theory as $\th \beta_{s} = \expa{ - 2 \beta_{g}}$ \cite{Kogut:79:1}. Thus in order to find the entropy of the probability distribution of  intersection points between electric strings and $\partial A$ we have performed Monte-Carlo simulations of the three-dimensional Ising model at the dual coupling, extracting the configurations of electric strings from the configurations of spin variables $z_{s}$ on some fixed time slice as described above. Since it is extremely difficult to measure exactly the $m$-point probability distributions $p_{m}\lr{x_{1}, \ldots, x_{m}}$, we have neglected the correlations between $x_{1}, \ldots, x_{m}$ and used the expression (\ref{entropy_simplification}).

 Lattice derivatives $S\lr{l+1} - S\lr{l}$ of the entanglement entropy and the entropy of the distribution of intersection points are plotted on Fig. \ref{fig:ment_vs_l} and Fig. \ref{fig:ment_vs_beta}. For the plot on Fig. \ref{fig:ment_vs_l} the coupling constants $\beta_{g} = 0.788$, $\beta_{s} = 0.21$ are close to the critical point with $\beta_{s \, c} \approx 0.22$, at which percolating electric strings emerge. It is reasonable to expect that for percolating strings the intersection points $\lrc{x_{1}, \ldots, x_{m}}$ are indeed uncorrelated. At these values of coupling constants the differences $S\lr{l+1} - S\lr{l}$ of both entropies indeed agree within the error range and practically do not depend on $l$, which means that the entropy is proportional to the area of the boundary with a good precision. We thus recover the familiar ``area law'' for the entanglement entropy \cite{Srednicki:93:1}. The data plotted on Fig. \ref{fig:ment_vs_beta} was obtained for the fixed size of the region $A$ $l = 8$ and for different values of the coupling constants $\beta_{g}$ and $\beta_{s}$. Again, only the differences $S\lr{l+1} - S\lr{l}$ were compared. The dependence on $\beta_{g}$ is similar for both entropies, however, in the strong coupling phase of the lattice gauge theory the entropy of intersection points is somewhat larger than the entanglement entropy. This discrepancy can be explained either by correlations between different intersection points or by some systematic error of the approximation used to calculate the entanglement entropy \cite{Buividovich:08:2}. It is not likely that the difference between the two entropies can be caused by additional entanglement between lattice degrees of freedom in the bulk of the regions $A$ and $B$, since in this case it is reasonable to expect that the entanglement entropy should be larger than the entropy on the boundary.

 It should be stressed that our results does not exclude the entanglement of states of electric strings which do not cross the boundary between $A$ and $B$, however, in this case the pattern of entanglement is the same as for the scalar field theories in the bulk of $A$ and $B$. Such situation may be realized when the entanglement entropy of gauge theories is calculated in the framework of AdS/CFT correspondence using the prescription of \cite{Ryu:06:1, Takayanagi:06:1, Klebanov:07:1}. In the quasiclassical approximation, in which the entanglement entropy is proportional to the minimal area of some hypersurface, the full string theory on AdS space is reduced to supergravity, which is the theory of particles rather than strings. Since the strings in AdS are believed to describe electric strings in the dual gauge theory, this means that the latter are very small and we are in the strong coupling limit (cf. the strong coupling limit of lattice gauge theory, where the dominant excitations are the electric strings which wind around one plaquette only). Correspondingly, there are very few strings that cross the boundary between $A$ and $B$, and the ``brane'' effect described above should be negligibly small.

 We thus see that the expression (\ref{trial_wf_ent_ent}) works well for all sizes of the region $A$ and for all values of the coupling constant, and the entanglement between the states in the bulk of the regions $A$ and $B$ seems to be rather small. The situation indeed looks as if the entanglement entropy was saturated by the uncertainty in the positions of the endpoints of open electric strings in the extended Hilbert space at the outer side of $\partial A$. In view of the actively discussed ``holographic descriptions'' of field theories, which relate the properties of string theories in the bulk of AdS space or modifications thereof with the properties of the field theories on its boundary \cite{tHooft:93:1, Susskind:95:1, Maldacena:97:1}, it seems rather tempting to associate the entanglement entropy of the region $A$ with the classical entropy of some statistical theory on its boundary. The endpoints of electric strings could then be interpreted as elementary degrees of freedom in such a theory. This can have interesting consequences in black hole physics - for instance, a confining gauge theory in the vicinity of a black hole might induce some two-dimensional field theory on its horizon, similarly to the scenario discussed in \cite{Carlip:99:1}. It could be also interesting to generalize the expression (\ref{trial_wf_ent_ent}) to the case of non-Abelian gauge theories, which is not a straightforward task, since the notion of conserved electric flux is not well understood for non-Abelian gauge groups.

\begin{acknowledgments}
 This work was partly supported by grants RFBR 06-02-04010-NNIO-a, RFBR 08-02-00661-a, DFG-RFBR 436 RUS, grant for scientific schools NSh-679.2008.2 and by Federal Program of the Russian Ministry of Industry, Science and Technology No 40.052.1.1.1112 and by Russian Federal Agency for Nuclear Power. The authors are grateful to E. Akhmedov, V. Shevchenko and V. Zakharov for interesting discussions on various aspects of entropy in field theories and gravity.
\end{acknowledgments}


\begin{thebibliography}{23}
\expandafter\ifx\csname natexlab\endcsname\relax\def\natexlab#1{#1}\fi
\expandafter\ifx\csname bibnamefont\endcsname\relax
  \def\bibnamefont#1{#1}\fi
\expandafter\ifx\csname bibfnamefont\endcsname\relax
  \def\bibfnamefont#1{#1}\fi
\expandafter\ifx\csname citenamefont\endcsname\relax
  \def\citenamefont#1{#1}\fi
\expandafter\ifx\csname url\endcsname\relax
  \def\url#1{\texttt{#1}}\fi
\expandafter\ifx\csname urlprefix\endcsname\relax\def\urlprefix{URL }\fi
\providecommand{\bibinfo}[2]{#2}
\providecommand{\eprint}[2][]{\url{#2}}

\bibitem[{\citenamefont{{'tHooft}}(1985)}]{tHooft:85:1}
\bibinfo{author}{\bibfnamefont{G.}~\bibnamefont{{'tHooft}}},
  \bibinfo{journal}{Nucl. Phys. B} \textbf{\bibinfo{volume}{256}},
  \bibinfo{pages}{727} (\bibinfo{year}{1985}).

\bibitem[{\citenamefont{Calabrese and Cardy}(2004)}]{Calabrese:04:1}
\bibinfo{author}{\bibfnamefont{P.}~\bibnamefont{Calabrese}} \bibnamefont{and}
  \bibinfo{author}{\bibfnamefont{J.}~\bibnamefont{Cardy}}, \bibinfo{journal}{J.
  Stat. Mech.} \textbf{\bibinfo{volume}{0406}}, \bibinfo{pages}{002}
  (\bibinfo{year}{2004}), \urlprefix\url{http://arxiv.org/abs/hep-th/0405152}.

\bibitem[{\citenamefont{Calabrese and Cardy}(2006)}]{Calabrese:06:1}
\bibinfo{author}{\bibfnamefont{P.}~\bibnamefont{Calabrese}} \bibnamefont{and}
  \bibinfo{author}{\bibfnamefont{J.}~\bibnamefont{Cardy}},
  \bibinfo{journal}{Int.J.Quant.Inf.} \textbf{\bibinfo{volume}{4}},
  \bibinfo{pages}{429} (\bibinfo{year}{2006}),
  \urlprefix\url{http://arxiv.org/abs/quant-ph/0505193}.

\bibitem[{\citenamefont{Nishioka and Takayanagi}(2007)}]{Takayanagi:06:1}
\bibinfo{author}{\bibfnamefont{T.}~\bibnamefont{Nishioka}} \bibnamefont{and}
  \bibinfo{author}{\bibfnamefont{T.}~\bibnamefont{Takayanagi}},
  \bibinfo{journal}{JHEP} \textbf{\bibinfo{volume}{01}}, \bibinfo{pages}{090}
  (\bibinfo{year}{2007}), \urlprefix\url{http://arxiv.org/abs/hep-th/0611035}.

\bibitem[{\citenamefont{Klebanov et~al.}(2007)\citenamefont{Klebanov, Kutasov,
  and Murugan}}]{Klebanov:07:1}
\bibinfo{author}{\bibfnamefont{I.~R.} \bibnamefont{Klebanov}},
  \bibinfo{author}{\bibfnamefont{D.}~\bibnamefont{Kutasov}}, \bibnamefont{and}
  \bibinfo{author}{\bibfnamefont{A.}~\bibnamefont{Murugan}},
  \emph{\bibinfo{title}{Entanglement as a probe of confinement}}
  (\bibinfo{year}{2007}), \urlprefix\url{http://arxiv.org/abs/0709.2140}.

\bibitem[{\citenamefont{Velytsky}(2008)}]{Velytsky:08:1}
\bibinfo{author}{\bibfnamefont{A.}~\bibnamefont{Velytsky}},
  \bibinfo{journal}{Phys. Rev. D} \textbf{\bibinfo{volume}{77}},
  \bibinfo{pages}{085021} (\bibinfo{year}{2008}),
  \urlprefix\url{http://arxiv.org/abs/0801.4111}.

\bibitem[{\citenamefont{Buividovich and Polikarpov}(2008)}]{Buividovich:08:2}
\bibinfo{author}{\bibfnamefont{P.~V.} \bibnamefont{Buividovich}}
  \bibnamefont{and} \bibinfo{author}{\bibfnamefont{M.~I.}
  \bibnamefont{Polikarpov}}, \emph{\bibinfo{title}{Numerical study of
  entanglement entropy in {SU(2)} lattice gauge theory}}
  (\bibinfo{year}{2008}), \urlprefix\url{http://arxiv.org/abs/0802.4247}.

\bibitem[{\citenamefont{Bennett et~al.}(1996)\citenamefont{Bennett, Bernstein,
  Popescu, and Schumacher}}]{Bernstein:96:1}
\bibinfo{author}{\bibfnamefont{C.~H.} \bibnamefont{Bennett}},
  \bibinfo{author}{\bibfnamefont{H.~J.} \bibnamefont{Bernstein}},
  \bibinfo{author}{\bibfnamefont{S.}~\bibnamefont{Popescu}}, \bibnamefont{and}
  \bibinfo{author}{\bibfnamefont{B.}~\bibnamefont{Schumacher}},
  \bibinfo{journal}{Phys. Rev. A} \textbf{\bibinfo{volume}{53}},
  \bibinfo{pages}{2046 } (\bibinfo{year}{1996}),
  \urlprefix\url{http://link.aps.org/abstract/PRA/v53/p2046}.

\bibitem[{\citenamefont{Ryu and Takayanagi}(2006)}]{Ryu:06:1}
\bibinfo{author}{\bibfnamefont{S.}~\bibnamefont{Ryu}} \bibnamefont{and}
  \bibinfo{author}{\bibfnamefont{T.}~\bibnamefont{Takayanagi}},
  \bibinfo{journal}{Phys. Rev. Lett.} \textbf{\bibinfo{volume}{96}},
  \bibinfo{pages}{181602} (\bibinfo{year}{2006}),
  \urlprefix\url{http://arxiv.org/abs/hep-th/0603001}.

\bibitem[{\citenamefont{Polyakov}(1987)}]{PolyakovGaugeStrings}
\bibinfo{author}{\bibfnamefont{A.~M.} \bibnamefont{Polyakov}},
  \emph{\bibinfo{title}{Gauge Fields and Strings}} (\bibinfo{publisher}{Harwood
  Academic Publishers}, \bibinfo{year}{1987}).

\bibitem[{\citenamefont{Kogut and Susskind}(1975)}]{Kogut:75:1}
\bibinfo{author}{\bibfnamefont{J.}~\bibnamefont{Kogut}} \bibnamefont{and}
  \bibinfo{author}{\bibfnamefont{L.}~\bibnamefont{Susskind}},
  \bibinfo{journal}{Phys. Rev. D} \textbf{\bibinfo{volume}{11}},
  \bibinfo{pages}{395} (\bibinfo{year}{1975}),
  \urlprefix\url{http://prola.aps.org/abstract/PRD/v11/i2/p395}.

\bibitem[{\citenamefont{Kogut}(1979)}]{Kogut:79:1}
\bibinfo{author}{\bibfnamefont{J.}~\bibnamefont{Kogut}}, \bibinfo{journal}{Rev.
  Mod. Phys.} \textbf{\bibinfo{volume}{51}}, \bibinfo{pages}{659 }
  (\bibinfo{year}{1979}),
  \urlprefix\url{http://prola.aps.org/abstract/RMP/v51/i4/p659_1}.

\bibitem[{\citenamefont{Hamma et~al.}(2005)\citenamefont{Hamma, Ionicioiu, and
  Zanardi}}]{Hamma:05:1}
\bibinfo{author}{\bibfnamefont{A.}~\bibnamefont{Hamma}},
  \bibinfo{author}{\bibfnamefont{R.}~\bibnamefont{Ionicioiu}},
  \bibnamefont{and} \bibinfo{author}{\bibfnamefont{P.}~\bibnamefont{Zanardi}},
  \bibinfo{journal}{Phys. Rev. A} \textbf{\bibinfo{volume}{71}},
  \bibinfo{pages}{022315} (\bibinfo{year}{2005}),
  \urlprefix\url{http://arxiv.org/abs/quant-ph/0409073}.

\bibitem[{\citenamefont{Levin and Wen}(2006)}]{Levin:06:1}
\bibinfo{author}{\bibfnamefont{M.}~\bibnamefont{Levin}} \bibnamefont{and}
  \bibinfo{author}{\bibfnamefont{X.}~\bibnamefont{Wen}},
  \bibinfo{journal}{Phys. Rev. Lett.} \textbf{\bibinfo{volume}{96}},
  \bibinfo{pages}{110405} (\bibinfo{year}{2006}),
  \urlprefix\url{http://arxiv.org/abs/cond-mat/0510613}.

\bibitem[{\citenamefont{Donnelly}(2008)}]{Donnelly:08:1}
\bibinfo{author}{\bibfnamefont{W.}~\bibnamefont{Donnelly}},
  \emph{\bibinfo{title}{Entanglement entropy in loop quantum gravity}}
  (\bibinfo{year}{2008}), \urlprefix\url{http://arxiv.org/abs/0802.0880}.

\bibitem[{\citenamefont{Bombelli et~al.}(1986)\citenamefont{Bombelli, Rabinder,
  Lee, and Sorkin}}]{Bombelli:86:1}
\bibinfo{author}{\bibfnamefont{L.}~\bibnamefont{Bombelli}},
  \bibinfo{author}{\bibfnamefont{K.~K.} \bibnamefont{Rabinder}},
  \bibinfo{author}{\bibfnamefont{J.}~\bibnamefont{Lee}}, \bibnamefont{and}
  \bibinfo{author}{\bibfnamefont{R.~D.} \bibnamefont{Sorkin}},
  \bibinfo{journal}{Phys. Rev. D} \textbf{\bibinfo{volume}{34}},
  \bibinfo{pages}{373} (\bibinfo{year}{1986}),
  \urlprefix\url{http://prola.aps.org/abstract/PRD/v34/i2/p373}.

\bibitem[{\citenamefont{Fursaev}(2006)}]{Fursaev:06:1}
\bibinfo{author}{\bibfnamefont{D.~V.} \bibnamefont{Fursaev}},
  \bibinfo{journal}{Phys. Rev. D} \textbf{\bibinfo{volume}{73}},
  \bibinfo{pages}{124025} (\bibinfo{year}{2006}),
  \urlprefix\url{http://arxiv.org/abs/hep-th/0602134}.

\bibitem[{\citenamefont{Greensite}(1979)}]{Greensite:79:1}
\bibinfo{author}{\bibfnamefont{J.~P.} \bibnamefont{Greensite}},
  \bibinfo{journal}{Nucl. Phys. B} \textbf{\bibinfo{volume}{158}},
  \bibinfo{pages}{469 } (\bibinfo{year}{1979}).

\bibitem[{\citenamefont{Srednicki}(1993)}]{Srednicki:93:1}
\bibinfo{author}{\bibfnamefont{M.}~\bibnamefont{Srednicki}},
  \bibinfo{journal}{Phys. Rev. Lett.} \textbf{\bibinfo{volume}{71}},
  \bibinfo{pages}{666} (\bibinfo{year}{1993}),
  \urlprefix\url{http://arxiv.org/abs/hep-th/9303048}.

\bibitem[{\citenamefont{{t'Hooft}}()}]{tHooft:93:1}
\bibinfo{author}{\bibfnamefont{G.}~\bibnamefont{{t'Hooft}}},
  \emph{\bibinfo{title}{Dimensional reduction in quantum gravity}},
  \urlprefix\url{http://arxiv.org/abs/hep-th/9409089}.

\bibitem[{\citenamefont{Susskind}(1995)}]{Susskind:95:1}
\bibinfo{author}{\bibfnamefont{L.}~\bibnamefont{Susskind}},
  \bibinfo{journal}{J. Math. Phys.} \textbf{\bibinfo{volume}{36}},
  \bibinfo{pages}{6377} (\bibinfo{year}{1995}),
  \urlprefix\url{http://arxiv.org/abs/hep-th/9409089}.

\bibitem[{\citenamefont{Maldacena}(1997)}]{Maldacena:97:1}
\bibinfo{author}{\bibfnamefont{J.~M.} \bibnamefont{Maldacena}},
  \bibinfo{journal}{Int.J.Theor.Phys.} \textbf{\bibinfo{volume}{38}},
  \bibinfo{pages}{1113} (\bibinfo{year}{1997}),
  \urlprefix\url{http://arxiv.org/abs/hep-th/9711200}.

\bibitem[{\citenamefont{Carlip}(1999)}]{Carlip:99:1}
\bibinfo{author}{\bibfnamefont{S.}~\bibnamefont{Carlip}},
  \bibinfo{journal}{Phys. Rev. Lett.} \textbf{\bibinfo{volume}{82}},
  \bibinfo{pages}{2828} (\bibinfo{year}{1999}),
  \urlprefix\url{http://arxiv.org/abs/hep-th/9812013}.

\end{thebibliography}

\end{document}